\documentclass[12pt]{article}

\usepackage{graphicx}
\usepackage{cite}
\usepackage{bm}
\usepackage{preprint2}

\newcommand\JCCvec[1]{\boldsymbol{\mathbf #1}}
\newcommand\trans[1]{{\JCCvec{#1}}_T}

\begin{document}

\preprint{hep-ph/0204004}

\title{Leading-twist Single-transverse-spin asymmetries: Drell-Yan and
  Deep-Inelastic Scattering}

\author{John C. Collins\thanks{
        E-mail: {\tt collins@phys.psu.edu}}
    \\
        Physics Department, Pennsylvania State University\\
        104 Davey Lab., University Park, PA 16802-6300,
        U.S.A.
}

\date{May 2, 2002}

\maketitle
\thispagestyle{empty}

\begin{abstract}
  Recently, Brodsky, Hwang and Schmidt have proposed a new mechanism
  that gives a transverse spin symmetry at leading twist in
  semi-inclusive deep-inelastic scattering.  I show that the new
  mechanism is compatible with factorization and is due to an
  transverse-spin asymmetry in the $k_T$ distribution of quarks in a
  hadron (the ``Sivers asymmetry'').  An earlier proof that the Sivers
  asymmetry vanishes because of time-reversal invariance is
  invalidated by the path-ordered exponential of the gluon field in
  the operator definition of parton densities.  Instead, the
  time-reversal argument shows that the Sivers asymmetry is reversed
  in sign in hadron-induced hard processes (e.g., Drell-Yan), thereby
  violating naive universality of parton densities.  Previous
  phenomenology with time-reversal-odd parton densities is therefore
  validated.
\end{abstract}

%============================================
\section{Introduction}
\label{sec:intro}

Recently, Brodsky, Hwang and Schmidt \cite{BHS} (BHS) have
demonstrated the existence of a new leading-twist mechanism for a
single-spin asymmetry in semi-inclusive deep-inelastic scattering
(SIDIS).  This is an important result for two reasons.  First is that
a single-spin asymmetry (SSA) of this kind directly probes the
partonic structure associated with chiral-symmetry breaking.  Second
is the paucity of leading-twist observables in hard-scattering
processes with an SSA.  Barone, Drago and Ratcliffe \cite{review} have
recently published a useful review of the subject.

In this letter, I will explore some important consequences of the
results of BHS:
\begin{itemize}
\item It is completely consistent with factorization, and provides an
  existence proof of the Sivers asymmetry \cite{Sivers}, that is, of a
  transverse-spin-dependent azimuthal asymmetry of the distribution of
  quarks in a proton.
\item Although this asymmetry is time-reversal odd, it is nonzero.  A
  proof \cite{JCC.one} that I gave to the contrary must therefore be
  incorrect.
\item Hence phenomenology that has been done using the Sivers asymmetry
  \cite{use.Sivers,T.odd.RHIC} is in fact appropriate in QCD.  
\item Instead the time-reversal argument shows that the asymmetry is
  reversed in the Drell-Yan process.
\item Hence there is an SSA in the Drell-Yan process.  This is of course of
  direct importance \cite{T.odd.RHIC} to the RHIC spin program.
\item Another time-reversal odd parton distribution is permitted
  \cite{review,classification,T.odd.DIS}: this is labeled $h_1^\perp$,
  and it gives a transversity to quarks taken from an {\em
    un}polarized hadron. (The transversity is dependent on the
  transverse momentum of the quark.)
\item Hence in the Drell-Yan process, even when both the beam and
  target are unpolarized, the annihilating quark and antiquark have a
  transverse-momentum-dependent transversity.  As shown by
  Brandenburg, Nachtmann and Mirkes \cite{explain.DY}, this gives a
  characteristic angular dependence for the lepton pair pair, a term
  proportional to $\sin^2\theta\cos2\phi$.  Such an asymmetry has been
  observed \cite{DY.azimuth} experimentally, and its large size has
  been a phenomenological puzzle.  The only explanation
  \cite{explain.DY,T.odd.RHIC} has been a nonzero $h_1^\perp$ parton
  density, which has previously appeared to be ruled out by
  time-reversal invariance.  Boer \cite{T.odd.RHIC} has both fitted
  the data and discussed its implications for the RHIC spin program.
\end{itemize}
The phenomenology of the various parton densities will of course be
rather complicated to sort out \cite{T.odd.DIS}, since in addition to
the two mechanisms listed above, there is also the asymmetry
\cite{JCC.one} associated with the fragmentation of transversely
polarized quarks.  The different mechanisms can be distinguished by
their different dependences on the transverse spin of the proton and
on the azimuthal direction relative to the outgoing lepton.  In
addition to the Drell-Yan data, there is also data \cite{HERMES} on
the SSA in SIDIS, which up to now has been only analyzed in terms of
the fragmentation effect.  So one knows the effects are quite
substantial.

The physics importance of these analyses is that the parton densities
involved arise from interference between amplitudes with left- and
right-handed polarization states, so that they only exist because of
chiral symmetry breaking in (non-perturbative) QCD.  They therefore
provide new probes of the chiral nature of the partonic structure of
hadrons, as well as of the interactions that produce the necessary
phases.

%===========================================================
\section{Deep-inelastic scattering}
\label{sec:DIS}

BHS performed their calculations in a field-theoretic model in which
QCD, with massive quarks, is supplemented by a colored scalar diquark
field and an elementary proton field.  Independently of whether this
model is suitable for describing nonperturbative hadronic physics, it
provides an excellent testbed for matters of principle, for example
whether an SSA is permitted by the symmetries of QCD.  The quark and
proton are massive, so chiral symmetry is broken, just as in QCD.  If
the gauge field is abelian, the gluon can be given a mass in the
Lagrangian without violating renormalizability, and then the
calculation can be done without confusion by the effects of actual
soft or collinear divergences.  This takes us further from real QCD,
but leaves the matters of principle unaffected.

\begin{figure}
   \centering
   \includegraphics[scale=0.45]{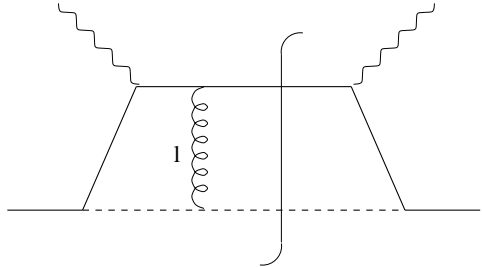}
   \caption{Graph with SSA for SIDIS.}
   \label{fig:SSA.SIDIS}
\end{figure}

As is well-known, the existence of an SSA requires a non-trivial phase
in the amplitude, and since the initial hadronic state is a single
stable particle, the phase is associated with final state
interactions.  In the lowest-order graphs, Fig.\ \ref{fig:SSA.SIDIS}
and its hermitian conjugate, the SSA therefore arises from the
imaginary part of the amplitude, which can be calculated by setting
the intermediate quark-diquark state on-shell.

By explicit calculation Brodsky, Hwang and Schmidt \cite{BHS} have
shown that there is indeed a leading-twist SSA.  The effect is only
large if the transverse momentum of the outgoing particle is low
compared with $Q$.  Otherwise a standard factorization theorem without
partonic transverse momentum would be appropriate, and then the SSA
would be power suppressed (of ``twist-3'').  Of course, most of the
cross section is at low transverse momentum.  Any relevant
factorization theorem will involve transverse-momentum-dependent
parton distribution functions (pdfs) and fragmentation functions, as
opposed to the more commonly used integrated distributions.

Since the momentum $l^\mu$ of the exchanged gluon is in a region
associated with rescattering, the calculated SSA appears, at first
sight, to be in contradiction with the factorization theorem, and its
proof.  But a closer examination of a proof of factorization shows
that this first impression is false.  As explained\footnote{ The
  arguments in Ref.\ \cite{diff.fact} are tailored to diffractive DIS,
  but they apply quite generally.  } in Ref.\ \cite{diff.fact} the
integration over $l^\mu$ must be deformed far into the complex plane in
order to derive factorization; the contour is not pinched in the
rescattering region.  The quark line may then be replaced by an
eikonalized line, which in turn can be obtained from the Feynman rules
for the Wilson line in the gauge-invariant operator definition of the
quark number density --- see Eq.\ (\ref{eq:pdf.def}) below.  The SSA
in the cross section is therefore due to an SSA in the dependence of
the parton density on transverse momentum.

Because the contour can be deformed, the quark propagator is in fact
off-shell by order $Q^2$, and the final-state interactions occur over
a short scale in time and distance, the same scale as is associated
with the uncertainty in the position of the vertex of the virtual
photon.  Hence the final-state interactions are definitely {\em not}
those associated with hadronization of the quark.

The parton density has the following operator definition \cite{CS}
\begin{eqnarray}
  \label{eq:pdf.def}
  P(x, \trans{k}, \trans{s}, \zeta)
  &=& \int \frac{dy^- \, d^2\trans{y} }{(2\pi)^3}
    e^{-ixp^+y^- + i\trans{k}\cdot\trans{y}}
\times 
\nonumber\\&& ~ ~ ~
\times
   \langle p| \bar\psi(0,y^-,\trans{y}) W_{y\infty}^{\dag}
     \frac{\gamma^+}{2}
     W_{0\infty} \psi(0) |p\rangle.
\end{eqnarray}
Here, light-front coordinates are used: $y^\mu = (y^+,y^-,\trans{y}) = (
(y^0 \pm y^z)/\sqrt2, \trans{y})$.  As usual, $x$ and $\trans{k}$ are the
fractional longitudinal momentum and the transverse momentum of the quark,
and $\trans{s}$ is the transverse part of the proton's spin
vector\footnote{The number density may depend on the proton's transverse
  spin, but parity invariance prohibits a dependence on the proton's
  helicity.  }.  The symbol $W_{y\infty}$ indicates a Wilson-line operator
going out from the point $y$ to future infinity, and $\zeta$ is
a variable associated with its direction, as we will now explain.

The form of the eikonal approximation for an outgoing quark determines
the future as the correct direction, and suggests that the Wilson
lines go in the light-like direction given by the vector $n^\mu =
(n^+,n^-,\trans{n}) = (0,1,\trans{0})$.  However, as shown by Collins
and Soper \cite{b.to.b}, the use of a light-like direction gives
severe divergences associated with integrals over the rapidity of
gluons.  It is suitable to cutoff the divergences by choosing a
non-light-like direction \cite{Sudakov}, parameterized by the variable
$\zeta$ in Eq.\ (\ref{eq:pdf.def}).  Solution of the equation
\cite{b.to.b} for evolution in $\zeta$ gives well-known Sudakov effects
that broaden the $k_T$ distribution; Meng, Olness and Soper
\cite{SIDIS.lowqt} have investigated its phenomenology in SIDIS.
Nevertheless this complication does not affect the lowest-order
calculation of the SSA.

The parton number density defined in Eq.\ (\ref{eq:pdf.def}) can be
decomposed \cite{classification,T.odd.DIS} into the conventional
spin-independent term and a spin-dependence $\trans{s} \times \trans{k}/M$
times a special parton density, labeled $f^\perp_{1T}$.

\begin{figure}
   \centering
   \includegraphics[scale=0.45]{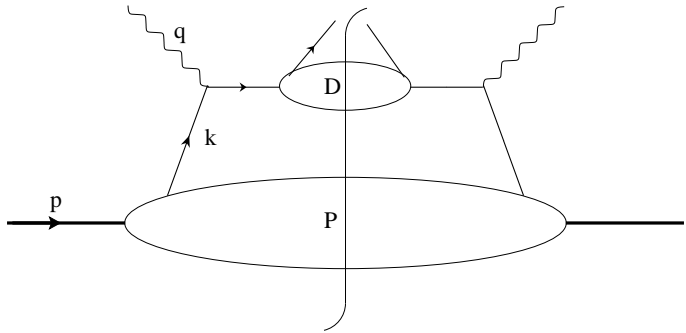}
   \caption{Factorization at low transverse momentum for SIDIS.}
   \label{fig:fact}
\end{figure}

Calculations can be readily performed to reproduce the BHS result from the
SSA of the parton density defined by Eq,\ (\ref{eq:pdf.def}).  The
factorization theorem to which it is associated is illustrated in Fig.\ 
\ref{fig:fact}, where there are transverse-momentum-dependent quark
distribution and fragmentation functions.

%=============================================
\section{Why the Sivers asymmetry is allowed}

This SSA in the quark density is exactly the one proposed by Sivers
\cite{Sivers}.  Now, in \cite{JCC.one} I gave a proof that the Sivers
asymmetry vanishes.  The proof involved applying space- and
time-reversal to the quark fields in the operator definition of the
parton densities, and the proof fails because it ignored the presence
of Wilson lines in the operators defining the parton density.  Under
time-reversal the future-pointing Wilson lines are replaced by
past-pointing Wilson lines, so that the correct version of the proof
gives
\begin{equation}
  \label{eq:PT}
  P(x, \trans{k}, \trans{s}, \zeta) |_{{\rm future-pointing}~W}
  = P(x, \trans{k}, -\trans{s}, \zeta) |_{{\rm past-pointing}~W} .
\end{equation}
Observe the change in sign of the transversity vector.  Since, as we will
see, the past-pointing Wilson lines are appropriate for factorization in
the Drell-Yan process, the correct result is not that the Sivers asymmetry
vanishes, but that it has opposite signs in DIS and in Drell-Yan:
\begin{equation}
  \label{eq:PT1}
  f^\perp_{1T}(x,k_T,\zeta)|_{\rm DIS} = -f^\perp_{1T}(x,k_T,\zeta)|_{\rm DY} .
\end{equation}

In a sense, I have derived the situation suggested by Anselmino,
Barone, Drago and Murgia \cite{ABDM}, that non-standard time-reversal
properties can enable the Sivers asymmetry to exist.  For the
elementary parton fields, the standard time-reversal transformation is
a symmetry of the Lagrangian, and we cannot evade its consequences.
But the fields defining the parton densities are non-local: they are
the elementary fields multiplied by Wilson lines.  Perhaps these
non-local fields can be related to the chiral quark fields discussed
in \cite{ABDM}.

In the approximation that the Wilson lines are along light-like lines,
the Wilson lines in the operator definition of the parton densities
can be eliminated by using the light-cone gauge.  However, the gluon
propagator then contains singularities of the form $1/k^+$.  It
follows from the above discussion that the prescription for defining
the analytic properties of this singularity must be intimately tied
with the derivation of the factorization theorem.  Therefore a simple
use of standard prescriptions, such as those of Refs.\
\cite{LB,Leibbrandt,Kovchegov}, is likely to be incorrect.

BHS argue that defining the parton densities by the light-cone gauge
method immediately relates them to the exclusive light-cone wave
functions.  Then they argue that since these wave functions are
properties of the proton state alone, they cannot involve final-state
interactions and cannot generate the SSA under discussion.  This
appears to be in contradiction with the arguments given above.

\begin{figure}
   \centering
   \includegraphics[scale=0.5]{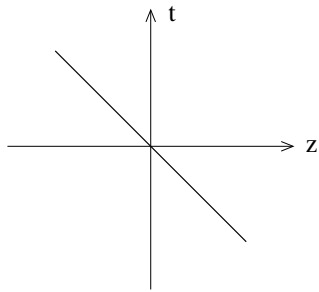}
   \caption{Light-like surface used to define the light-front wave
      functions and parton densities. }
   \label{fig:space-time}
\end{figure}

A resolution of this contradiction is that the light-cone wave
functions involve decomposing the proton state with respect to fields
on a light-like surface, as opposed to a space-like surface.  The
light-like eikonal line from which we obtained the phase giving the
SSA effectively gives the infinite-energy limit of the relevant
final-state interactions.  So the final-state interactions are
occurring on a light-like surface --- see Fig.\ \ref{fig:space-time}.
Final-state interactions can and do occur on a light-like surface, so
there need be no contradiction between saying that the SSA in the
parton density is given by final-state interactions and that the
parton density is obtained from wave-functions of the proton state.

Thus I would disagree with the claim ``Structure functions are not
parton probabilities'' that has been advanced by Brodsky et al.\
\cite{parton.probs}.  This issue deserves further study, of course.

%===========================================================
\section{Drell-Yan}
\label{sec:DY}

%-----------------------------
\subsection{SSA}

\begin{figure}
   \centering
   \includegraphics[scale=0.45]{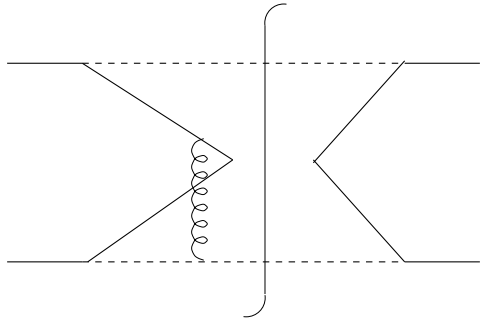}
   \caption{Graph with SSA for Drell-Yan.}
   \label{fig:SSA.DY}
\end{figure}

Let us now apply the BHS model to the Drell-Yan process, for which the
lowest-order graph giving an SSA is shown in Fig.\ \ref{fig:SSA.DY}.
The requisite imaginary part now comes from an {\em initial}-state
interaction between an active quark and a spectator diquark.\footnote{
  Interactions between the spectator diquarks alone cancel after a
  unitarity sum over final-states.  Observe that a corresponding
  unitarity sum is not possible in the DIS calculation, because the
  requirement of detecting a particular particle in the final state
  pins down the final-state cut.  In contrast, in fully inclusive DIS
  the unitarity sum is possible, and results in the well-known
  property that the $g_2$ structure function is of twist 3.  }

The contour deformation argument continues to apply, except that the
deformation is in the opposite direction, away from the initial-state
pole, and the Wilson line in the operator definition of the parton
density is therefore past-pointing.  As already explained, this
results in a reversed sign of the SSA as compared with SIDIS.

Close examination of the factorization proofs for Drell-Yan
\cite{CSS1,CSS2} will indicate that the past-pointing Wilson lines do
not mesh particularly well with later parts of the factorization proof
where cancellation of interactions between the two hadrons is
demonstrated.  These parts of the proof prefer future-pointing Wilson
lines.  However, the proofs were constructed in the particular context
of a cross section integrated over transverse momentum, or at large
transverse momentum.  Thus there is no automatic contradiction with
the new results.  However it is necessary to examine carefully how to
prove factorization in the case of the transverse-momentum-dependent
cross sections.  As regards transverse-momentum distributions, the
actual proofs of Collins and Soper \cite{b.to.b} only apply to
$e^+e^-$ annihilation, a process with only final-state interactions.

%----------------------------
\subsection{Unpolarized Drell-Yan}

The failure of the time-reversal argument applies to more than the Sivers
asymmetry.  There is also the possibility of a nonzero transversity of a
quark in an unpolarized hadron, with the transversity being correlated
with the transverse momentum of the quark.  This is defined by
\cite{review,classification,T.odd.DIS}
\begin{eqnarray}
  \label{eq:pdf2.def}
  \frac{k^j}{M}\epsilon_{ij} h_1^\perp(x, k_T^2, \zeta)
  &=& \int \frac{dy^- \, d^2\trans{y} }{(2\pi)^3}
    e^{-ixp^+y^- + i\trans{k}\cdot\trans{y}}
\nonumber\\&& ~ ~ ~
   \langle p| \bar\psi(0,y^-,\trans{y}) W_{y,-\infty}^{\dag}
     \frac{\gamma^+}{2} \gamma_i \gamma_5
     W_{0,-\infty} \psi(0) |p\rangle.
\end{eqnarray}
In the BHS model, the value of this object ought to be very similar to the
previously discussed SSA, since the calculation is quite similar.

Among the phenomenological implications of this quark transversity is
an important result for the angular distribution of the final-state
leptons in the Drell-Yan process.  Since the transversities of the
annihilating quark and antiquark are correlated with the transverse
momentum of the Drell-Yan pair, there is a characteristic angular
dependence for the leptons, from the spin-dependence of the elementary
$q\bar{q}\to l^+l^-$ process.  The angular dependence \cite{explain.DY}
is proportional to $\sin^2\theta\cos2\phi$.

Now it has long been known that experimental measurements
\cite{DY.azimuth} of the angular distribution of the leptons have a
coefficient for this part of the angular dependence that is
substantially larger than that predicted by the hard-scattering
calculations for Drell-Yan pairs of {\em large} transverse momentum.
Moreover, Brandenburg, Nachtmann and Mirkes \cite{explain.DY} have
shown that this large coefficient is explained by a substantial
non-zero value for the $h_1^\perp$ quark density; a more recent fit was
given by Boer \cite{T.odd.RHIC}.  The apparent contradiction, that
this quark density is prohibited by time-reversal invariance of QCD
has now disappeared.

%=============================================
\section{Conclusions}

We now see that there are three sources for the SSA and related
azimuthal dependences in leading-twist processes:
\begin{itemize}
\item The Sivers asymmetry \cite{Sivers} with transversely polarized
  target.
\item The final-state fragmentation asymmetry from a transversely
  polarized quark, with the quark acquiring its transversity from the
  initial-state transverse polarization of the target proton.  This
  includes both the single-particle asymmetry of Ref.\ \cite{JCC.one}
  and the two-particle asymmetry of Ref.\ \cite{other.frag} and
  \cite{JJT}
\item The transversity of a quark in an unpolarized hadron
  \cite{review,classification,T.odd.DIS}. 
\end{itemize}
Previously, only the fragmentation asymmetries appeared to be
permitted by time-reversal invariance.  But now we have seen that the
presence of Wilson lines in the definition of the parton densities
also allows the time-reversal-odd parton densities to be nonzero.

This of course uncovers a rich phenomenology.  Luckily a start has
made, notably by Anselmino and Murgia \cite{use.Sivers}, by Boer
\cite{T.odd.RHIC}, and by Boer and Mulders \cite{T.odd.DIS}, who have
chosen to follow experimental indications of time-reversal-odd parton
densities.  In particular, the measured angular dependence of leptons
\cite{DY.azimuth} in the unpolarized Drell-Yan process indicates that
substantial effects exist.

Hence new areas of research are possible at RHIC, for example with the
Drell-Yan asymmetry.  Comparison of results between DIS and Drell-Yan
will be particularly interesting.  A test of the prediction of the
reversal of the sign of the Sivers asymmetry between the two processes
is absolutely vital, since it would validate the whole approach of
this paper.

In previous work on factorization, the issue of the definition of the
Wilson lines has appeared to be a technicality.  But now the sign
difference between DIS and Drell-Yan provides an experimental probe
sensitive to this issue and hence to the time-dependence of the
associated microscopic physics.

One complication that it will be essential to treat correctly is the
Sudakov driven dilution of the asymmetries in
transverse-momentum-dependent cross sections as the overall energy and $Q$
are increased.

The results of all this work will be measurements of parton-level
asymmetries that probe the dynamics of chiral-symmetry breaking, and
that can therefore connect in an interesting way to theories of
hadronic structure.

%============================================
\section*{Notes added}

An earlier account of the phenomenology of the Sivers effect than
\cite{use.Sivers} was given by Anselmino, Boglione and Murgia in Ref.\ 
\cite{use.Sivers2}.

See \cite{HERMES.analysis} for analyses of the HERMES data
\cite{HERMES}. There are earlier discussions \cite{T.odd.DIS,BM} that
include the Sivers asymmetry in this context.

At least one additional related effect\footnote{I thank M. Anselmino
  for pointing this out.}  can be used in this area: the polarization
of a $\Lambda$ baryon in the fragmentation of an unpolarized quark, as a
function of their relative transverse momentum --- see Ref.\ 
\cite{Lambda}.

%============================================
\section*{Acknowledgments}

This work was supported in part by the U.S. Department of Energy under
grant number DE-FG02-90ER-40577.  I would like to thank Stan Brodsky,
Steve Heppelmann, Jianwei Qiu and Mark Strikman for useful
discussions.
I would also like to thank M. Anselmino, H. Avagyan, D.
Boer, and P. Mulders for comments on the first version of this paper.

%============================================

\end{document}